\documentclass[aps,showpacs,preprint,superscriptaddress]{revtex4}
\usepackage{graphicx}
\usepackage{amsmath}
\usepackage{bbm}

\begin{document}

\title{Effects of electric field polarizations on pair production}

\author{Z. L. Li}
\author{D. Lu}
\affiliation{Key Laboratory of Beam Technology and Materials Modification of the Ministry of Education, College of Nuclear Science and Technology, Beijing Normal University, Beijing 100875, China}
\author{B. S. Xie\footnote{Corresponding author: bsxie@bnu.edu.cn}}
\affiliation{Key Laboratory of Beam Technology and Materials Modification of the Ministry of Education, College of Nuclear Science and Technology, Beijing Normal University, Beijing 100875, China}
\affiliation{Beijing Radiation Center, Beijing 100875, China}

\date{\today}

\begin{abstract}
The effects of electric field polarizations on pair production from a vacuum are investigated numerically by employing the real-time Dirac-Heisenberg-Wigner formalism. For few-cycle fields, it is found that the interference pattern in momentum spectra is absent and the circular distortion of momentum distribution becomes more apparent with the increase of polarization. For multi-cycle fields, it is found that the interference effects in momentum spectra are obvious. And as the polarization increases, the momentum distribution is split into two parts in the momentum $q_y$ direction first and then two separated parts are connected into a ring. The effects of polarizations on the number density of created particles exhibits two different characteristics. For a small laser frequency, the particle number density deceases with the polarization, while for a large laser frequency, the relation between them is sensitive to the field frequency nonlinearly. Some interpretations for the obtained results in this study, which is expected to be valuable to deepen the understanding of pair production in complex fields and also helpful for the study associated to strong-field ionization.

\end{abstract}

\maketitle

\section{Introduction}

Electron-positron (EP) pair production from vacuum in the presence of a strong field is one of the well-known predictions of quantum electrodynamics (QED)  \cite{Sauter,Heisenberg,Schwinger} as a valuable probing the structure of QED vacuum. According to Schwinger's formula, the pair production rate ($\propto\exp(-\pi E_{\mathrm{cr}}/E)$) is exponentially suppressed for a weak field. So it seems that the experimental detection of vacuum pair creation is rather difficult because current electric field strength $E$ is far less than the critical electric field strength $E_{\mathrm{cr}}=m^2/e\sim1.32\times10^{16}\mathrm{V/cm}$ (the units $\hbar=c=1$ are used). However, the high-intensity and ultrashort laser facilities under construction, such as the Extreme Light Infrastructure \cite{ELI} and the x-ray free electron laser (XFEL) \cite{Ringwald,XFEL}, may provide a subcritical electric field which can produce a considerable number of EP pairs \cite{AlkoferPRL2001,RobertsPRL2002}. This stimulates the researcher renewed interests to explore significant phenomena of EP pair production from vacuum \cite{Dunne,Piazza1,Schutzhold,Abdukerim2013,Akkermans,Li2014}.

In recent years, many theoretical works have been done about the pair production in more complex and realistic fields. These studies can not only be available to the future experiment but also reveal many novel signatures which are useful for understanding the complicated physics in pair production. Among existed studies the polarization of fields is a key parameter. Let us recall some important results.

For a linearly polarized electric field, the momentum spectrum of created EP pairs in laser pulses with subcycle structures are investigated in Ref. \cite{Hebenstreit2009} with the help of quantum kinetic approach. It is found that the longitudinal momentum spectrum is extremely sensitive to the laser field parameters and its oscillations can be used as a probe of subcycle strucuturs in ultrashort laser pulses. Using the same electric field and calculation method, in Ref. \cite{Kohlfurst2014}, EP pair production in nonperturbative multiphoton regime are studied and the physical observable that carried the effective mass signatures were identified.
For a circularly polarized electric field that solely changes over time, Schwinger pair production is explored numerically with the real-time Dirac-Heisenberg-Wigner (DHW) formalism in Ref. \cite{Blinne2014}, and it is found that the characteristic momentum distribution in this field might be used to distinguish a QED cascade seeded by pair production from that seeded by isotropic vacuum impurity. Moreover, the pair creation rate for rotating electric fields is investigated semiclassically with the Wentzel-Kramers-Brillouin-type (WKB) approximation in Ref. \cite{Strobel2015}. It is found that for a certain range of electric field parameters the pair production rate was dominated by one spin of created pairs.

Let us turn to the case, i.e., the problem of EP pair production in a general elliptically polarized electric field. In fact we have performed some primary studies a few years ago. For example, we studied analytically the exponential factor of pair production by using the worldline instanton method \cite{Xie2012} and a two-level transition technique \cite{Mohamedsedik2012}, respectively. On the other hand, the elliptically polarized laser fields are also commonly used to explore new signatures of strong-field ionization (SFI) dynamics that can not be revealed by using a linearly polarized field \cite{PfeifferNatPhys2012,PfeifferPRL2012,HofmannJPB2013,HofmannPRA2014}. Due to the similarity between strong-field ionization and vacuum pair production, the study of EP pair creation in an elliptically polarized field can not only uncover some new phenomena of pair production but also be a reference for strong-field ionization. Furthermore, as a elliptically polarized laser field is easier to achieve experimentally than a perfect circularly polarized one, the former is more suitable for clarifying pair production mechanism as well as observing EP pair creation experimentally.

Motivated by the facts mentioned above that the importance of general elliptic polarization exists in both of EP pair creation and SFI, therefore, in this paper, we focus our study on the effects of electric field polarizations on EP pair production by numerically solving the real-time DHW formalism \cite{Bialynicki,Hebenstreit2010,Hebenstreit2011}. We calculate the momentum spectrum and the number density of created particles in different polarized electric fields with a same laser intensity and consider the effects of field polarizations on them. We find that with increasing polarizations the circular distortion of momentum distribution becomes more obvious for few-cycle fields while the momentum spectra are split for many-cycle fields. It is also found that the particle number density deceases with the increase of polarization for a small laser frequency and the relation between them is more complicated for a large laser frequency. These results are helpful for the understanding of pair creation in complex fields. Evenly the present study may also be used to build a bridge between vacuum pair creation and SFI.

This paper is organized as follows. In Sec. \ref{Theo-Form} we give a brief recall of the DHW formalism which is used in our calculation. The numerical results are presented and analyzed in Sec. \ref{Resu}. Section \ref{Conc-Diss} is the conclusion and discussion.

\section{Theoretical model and formulas}
\label{Theo-Form}

For a subcritical electric field $E\sim 0.1E_{\mathrm{cr}}$, it is a good approximation to neglect the collision effect and the internal electric field since the EP pair yield and the back-reaction electric current are quite small. And because the spatial scales of the EP pair production are smaller than the spatial focusing scales of the laser pulse, the spatial effects are not significant. Therefore, we have the spatially homogeneous and time-dependent fields. For our studies, we focus on the EP pair production in a uniform and time-varying electric field of arbitrary polarization
\begin{equation}\label{eq1}
\mathbf{E}(t)=\frac{E_0}{\sqrt{1+\delta^2}}\exp\Big(-\frac{t^2}{2\tau^2}\Big)\left[
                                             \begin{array}{c}
                                               \cos(\omega t+\phi) \\
                                               \delta\sin(\omega t+\phi) \\
                                               0 \\
                                             \end{array}
                                           \right],
\end{equation}
where $E_0$ is the maximal field strength, $\tau$ defines the pulse duration, $\omega$ is the laser frequency, $\phi$ is the carrier-envelope phase, and $ |\delta|\leq1$ represents the polarization (or the ellipticity). Note that a similar low-intensity electric field with the polarization up to $\pm0.93$ has been achieved experimentally \cite{PfeifferNatPhys2012}. For convenience, we set $\phi=0$ and $\delta>0$ unless otherwise specified.

Our following numerical results are based on the DHW formalism which has been used to study vacuum pair production in Refs. \cite{Bialynicki,Hebenstreit2010} for different electric fields. We start with the equal-time density operator of two Dirac field operators in the Heisenberg picture,
\begin{eqnarray}\label{eq2}
\hat{\mathcal{C}}_{\alpha\beta}(\mathbf{x},\mathbf{y},t)=&&e^{-ie\int^{1/2}_{-1/2}
\mathbf{A}(\mathbf{x}+\lambda \mathbf{y},t)\cdot \mathbf{y} d\lambda}\nonumber\\
&&\times\Big[\hat{\Psi}_\alpha\Big(\mathbf{x}+\frac{\mathbf{y}}{2},t\Big),
\hat{\bar{\Psi}}_\beta\Big(\mathbf{x}-\frac{\mathbf{y}}{2},t\Big)\Big],
\end{eqnarray}
with the center-of-mass coordinate $\mathbf{x}=(\mathbf{x}_1+\mathbf{x}_2)/2$ and the relative coordinate $\mathbf{y}=\mathbf{x}_1-\mathbf{x}_2$. Note that the factor before the commutator is a Wilson-line factor used to keep gauge invariance, and the integration path of the vector potential $\mathbf{A}$ is a straight line chosen to introduce a clearly defined kinetic momentum $\mathbf{p}$ . Moreover, we have employed a Hartree approximation for the electromagnetic field and chosen the temporal gauge $A_0=0$. The Wigner operator is defined as the Fourier transformation of Eq. (\ref{eq2}) with respect to the relative coordinate $\mathbf{y}$, and its vacuum expectation value gives the Wigner function
\begin{eqnarray}\label{eq3}
\mathcal{W}(\mathbf{x},\mathbf{p},t)=-\frac{1}{2}\int d^3ye^{-i\mathbf{p}\cdot\mathbf{y}}\langle0|\hat{\mathcal{C}}(\mathbf{x},\mathbf{y},t)|0\rangle.
\end{eqnarray}
Decomposing the Wigner function in terms of a complete basis set $\{\mathbbm{1},\gamma_5,\gamma^\mu,\gamma^\mu\gamma_5,\sigma^{\mu\nu}:=\frac{i}{2}
[\gamma^\mu,\gamma^\nu]\}$, we have
\begin{equation}\label{eq4}
\mathcal{W}(\mathbf{x},\mathbf{p},t)=\frac{1}{4}(\mathbbm{1}\mathbbm{s}
+i\gamma_5\mathbbm{p}+\gamma^\mu\mathbbm{v}_\mu
+\gamma^\mu\gamma_5\mathbbm{a}+\sigma^{\mu\nu}\mathbbm{t}_{\mu\nu}),
\end{equation}
with sixteen real Wigner components, scalar $\mathbbm{s}(\mathbf{x},\mathbf{p},t)$, pseudoscalar $\mathbbm{p}(\mathbf{x},\mathbf{p},t)$, vector $\mathbbm{v}(\mathbf{x},\mathbf{p},t)$, axialvector $\mathbbm{a}(\mathbf{x},\mathbf{p},t)$, and tensor $\mathbbm{t}(\mathbf{x},\mathbf{p},t)$. Inserting the decomposition into the equation of motion for the Wigner funtion, one can obtain a partial differential equation (PDE) system for the sixteen Wigner components \cite{Bialynicki}. Furthermore, for the spatially homogeneous and time-dependent electric fields mentioned above, by using the method of characteristics, or simply, replacing the kinetic momentum $\mathbf{p}$ by $\mathbf{q}-e\mathbf{A}(t)$ with the well-defined canonical momentum $\mathbf{q}$, the PDE system for the sixteen Wigner components can be reduced to an ordinary differential equation system for the ten nontrivial Wigner components $\mathbbm{w}(\mathbf{q},t)=(\mathbbm{s},\mathbbm{v},\mathbbm{a},\mathbbm{t}_1:
=2\mathbbm{t}^{i0}\mathbf{e}_i)^\textsf{T}(\mathbf{q},t)$,
\begin{equation}\label{eq5}
\dot{\mathbbm{w}}(\mathbf{q},t)=\mathcal{H}(\mathbf{q},t)\mathbbm{w}(\mathbf{q},t),
\end{equation}
where the dot denotes a total time derivative, $\mathcal{H}(\mathbf{q},t)$ is a $10\times10$ matrix.

The one-particle distribution function is defined as
\begin{equation}\label{eq6}
f(\mathbf{q},t)=\frac{1}{2}\mathbbm{e}^\textsf{T}_1\cdot[\mathbbm{w}(\mathbf{q},t)
-\mathbbm{w}_{\mathrm{vac}}(\mathbf{q},t)],
\end{equation}
where $\mathbbm{w}_{\mathrm{vac}}(\mathbf{q},t)=(\mathbbm{s}_{\mathrm{vac}}~\mathbbm{v}_{\mathrm{vac}}~ \mathbf{0} ~~\mathbf{0})^\textsf{T}$, $\mathbbm{s}_{\mathrm{vac}}=-2m/\Omega(\mathbf{p})|_{\mathbf{p}\rightarrow \mathbf{q}-e\mathbf{A}(t)}$, $\mathbbm{v}_{\mathrm{vac}}=-2\mathbf{p}/\Omega(\mathbf{p})|_{\mathbf{p}\rightarrow \mathbf{q}-e\mathbf{A}(t)}$, $\Omega(\mathbf{p})|_{\mathbf{p}\rightarrow \mathbf{q}-e\mathbf{A}(t)}=(m^2+[\mathbf{q}-e\mathbf{A}(t)]^2)^{1/2}$ is the total energy of electrons, and $\mathbbm{e}_1=-1/2~\mathbbm{w}_{\mathrm{vac}}$ is one of the basis of the ten-component vector $\mathbbm{w}$. Notice that the vacuum solution is $\mathbbm{w}_{\mathrm{vac}}(\mathbf{q},t_{\mathrm{vac}})$.

In order to precisely obtain the distribution function $f$, we employ the method used in \cite{Blinne2014}. Decomposing the Wigner components as  $\mathbbm{w}=2(f-1)\mathbbm{e}_1+\mathcal{F}\mathbbm{w}_9$ with an auxiliary nine-component vector $\mathbbm{w}_9$ and a $10\times9$ matrix
$\mathcal{F}=\left(
               \begin{array}{cccccc}
                 -\mathbf{p}^\textsf{T}/m ~~ \mathbf{0} \\
                 ~~~\mathbbm{1}_9 \\
               \end{array}
             \right)\Big|_{\mathbf{p}\rightarrow \mathbf{q}-e\mathbf{A}(t)},
$ and applying Eq. (\ref{eq5}), we have
\begin{eqnarray}\label{eq7}
\begin{array}{c}
  \dot{f}=1/2~\dot{\mathbbm{e}}_1^\textsf{T} \mathcal{F}\mathbbm{w}_9, \\
  \dot{\mathbbm{w}}_9=\mathcal{H}_9\mathbbm{w}_9+2(1-f)\mathcal{G}\dot{\mathbbm{e}}_1,
\end{array}
\end{eqnarray}
where $\mathcal{G}=(\mathbf{0}~~\mathbbm{1}_9)$ is a $9\times10$ matrix, and
\begin{equation}
\mathcal{H}_9=\left(
                \begin{array}{ccc}
                  -e\mathbf{p}\cdot \mathbf{E}^\textsf{T}/\omega^2(\mathbf{p}) & -2\mathbf{p}\times & -2m \\
                  -2\mathbf{p}\times & \mathbf{0} & \mathbf{0} \\
                  2(m^2+\mathbf{p}\cdot \mathbf{p}^\textsf{T})/m & \mathbf{0} & \mathbf{0} \\
                \end{array}
              \right)\bigg|_{\mathbf{p}\rightarrow \mathbf{q}-e\mathbf{A}(t)}.\nonumber
\end{equation}
Thus, we can get the one-particle momentum distribution function $f(\mathbf{q},t)$ by solving Eq. (\ref{eq7}) with the initial conditions $f(\mathbf{q},-\infty)=\mathbbm{w}_9(\mathbf{q},-\infty)=0$. Integrating the distribution function over full momenta at $t\rightarrow+\infty$, we have the number density of created pairs
\begin{equation}\label{eq8}
n(+\infty)=\int\frac{d^3q}{(2\pi)^3}f(\mathbf{q},+\infty).
\end{equation}

\section{Results}
\label{Resu}

Before analyzing our numerical results, we introduce the Keldysh adiabatic parameter $\gamma=m\omega/eE$ \cite{Keldysh} which is used to describe the dominant mechanism in pair production or strong-field ionization. For $\gamma\ll1$, the tunneling process dominate pair production and it is called nonperturbative Schwinger pair production. For $\gamma\gg1$, the main contribution to pair creation comes from multiphoton mechnism and this process is called perturbative multiphoton pair creation. However, when $\gamma\sim\mathcal{O}(1)$, multiphoton pair production will have a nonperturbative nature and many interesting features of pair production will be presented \cite{Kohlfurst2014,Ruf,Mocken}. This intermediate regime is also called nonperturbative multiphoton regime.

\begin{figure}[htbp]\suppressfloats
\includegraphics[width=8cm]{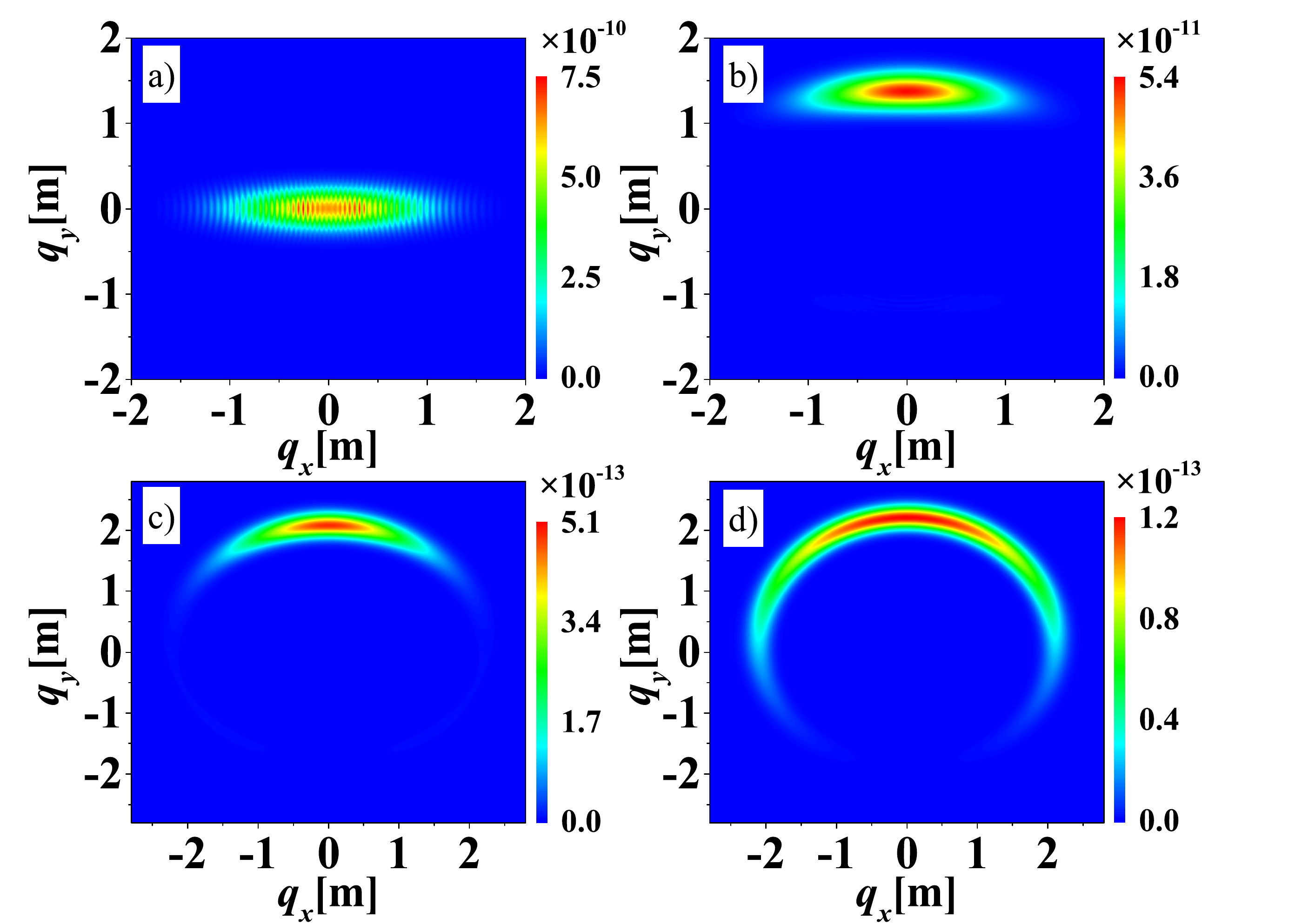}
\caption{\label{Fig1}(color online) Momentum spectra of created EP pairs in the $(q_x,q_y)$ plane for different polarizations with $\omega=0.05m$. (a) for $\delta=0$, (b) for $\delta=0.5$, (c) for $\delta=0.9$, and (d) for $\delta=1$. Other electric field parameters are chosen as $E_0=0.1\sqrt{2}E_{\mathrm{cr}}$ and $\tau=100/m$.}
\end{figure}

By solving Eq. (\ref{eq7}) numerically, we obtain the momentum spectra of created particles in the polarization plane $(q_x,q_y)$ for different polarizations with $\gamma=0.35\sqrt{1+\delta^2}$, see Fig. \ref{Fig1}. From Fig. \ref{Fig1}(a), one can see clearly the oscillatory structures of momentum spectrum. This result is similar to the one in \cite{Hebenstreit2009}, because for $\delta=0$ our electric field (\ref{eq1}) will become the linearly polarized field used in \cite{Hebenstreit2009}. When $\delta\neq0$, however, the oscillations vanish and the peak position of momentum distribution is shifted along the direction of $q_y$, see Fig. \ref{Fig1}(b). With the increase of polarization $\delta$, the momentum distribution of created EP pairs is distorted gradually. For $\delta\sim1$, it forms a ring-like structure. For this few-cycle field ($\sigma=\omega\tau=5$), the absence of interference effects in momentum distribution is because the polarization can easily change the distribution of tuning points (complex time $t_p$ satisfying $\Omega(\mathbf{q},t_p)=0$) and as a result only one pair of turning points dominates the pair creation. Moreover, since the electric field component $E_y(t)$ is an odd function of time, the momentum distribution is not symmetric about $q_y=0$ \cite{Dumlu2011}, this can also be seen in Fig. 4 of Ref. \cite{Hebenstreit2009}.

\begin{figure}[htbp]\suppressfloats
\includegraphics[width=12cm]{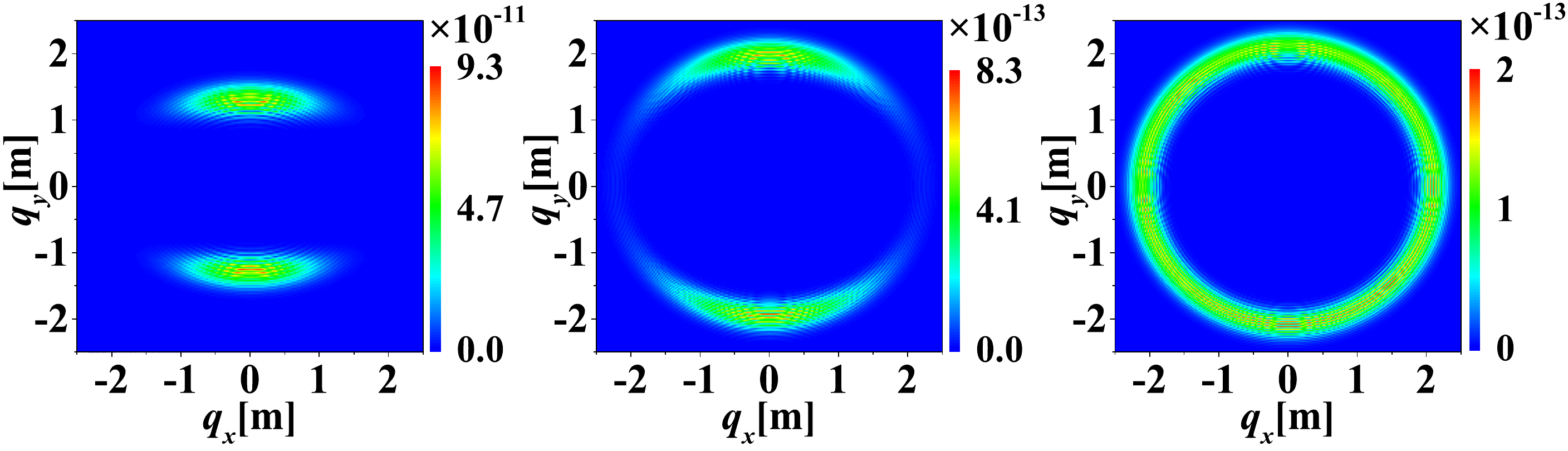}
\caption{\label{Fig2}(color online) Momentum spectra of created EP pairs in the $(q_x,q_y)$ plane for different polarizations with $\omega=0.05m$ and $\tau=300/m$. From left to right $\delta=0.5$, $0.9$, and $1$, respectively. The electric field strength $E_0=0.1\sqrt{2}E_{\mathrm{cr}}$.}
\end{figure}

Based on the analysis in Ref. \cite{Blinne2014}, it is known that for a circularly polarized field the circular distortion of momentum distribution can be enhanced by increasing the pulse duration, because the produced particles have enough time to follow the rotation of this field. And when the circularly distorted momentum distribution closes into a ring, interference effects will occur. An intuitively plausible understanding about this is that the quantum wave functions of left-moving and right-moving particles carrying different phases will superpose each other and induce the quantum interference. In fact, this effect can not be fully understood just by using circularly polarized fields. To solve this problem, we increase the cycle number of electric field ($\sigma=15$) by increasing the pulse duration and plot the momentum spectra of created pairs for different polarizations in Fig. \ref{Fig2}. Indeed, as discussed above, we can see a ring-like structure and the interference effect in the momentum spectrum for $\delta=1$, see Fig. \ref{Fig2}(c). Meanwhile, however, we also find that for a small value of $\delta$ the particle distribution in momentum space is split into two segments in the $q_y$ direction and interference effects exist in each segment. This result indicate clearly that the interference effect is related to the cycle number of electric field. Actually, it is caused by the interference among different pair-creation amplitudes corresponding to different turning points in the complex $t$ plane \cite{Strobel2015}. For a many-cycle field, there will be many pairs of turning points having nearly the same distance to the real $t$ axis and this can induce evident interference \cite{Dumlu2011}. Consequently, the interference effect in momentum spectrum will become more obvious with the cycle number of electric field increasing.

\begin{figure}[htbp]\suppressfloats
\includegraphics[width=15cm]{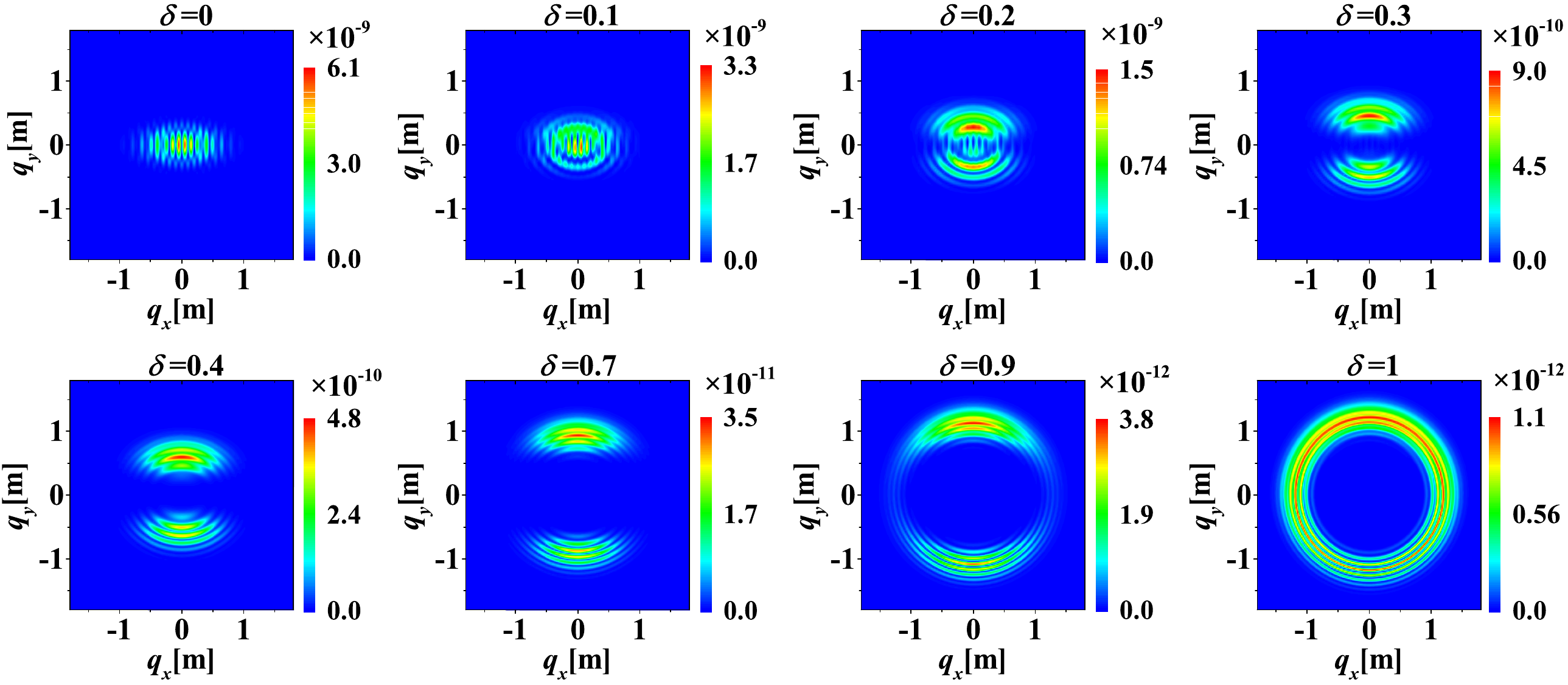}
\caption{\label{Fig3}(color online) Momentum spectra of created EP pairs in the $(q_x,q_y)$ plane for different polarizations with $\omega=0.1m$. The polarization value is shown in the headline of each panel. Other electric field parameters are the same as in Fig. \ref{Fig1}.}
\end{figure}

To see the effect of polarization on momentum spectra clearly, we show the momentum distribution of created pairs scanning over polarization with $\gamma=0.71\sqrt{1+\delta^2}$ in Fig. \ref{Fig3}. One can see that as the value of polarization increases the momentum distribution is split into two parts gradually in the $q_y$ direction, and then these two separated parts are connected to form a ring. More notably, when the value of polarization $\delta\gtrsim3$, the created particles with small momentum vanish. These findings are very similar to the results of strong-field ionization of helium using an elliptically polarized laser pulses (cf. Fig. 1 in Ref. \cite{PfeifferPRL2012}). Here we give an intuitive understanding about the split of momentum distribution shown in Fig. \ref{Fig3}. According to Schwinger's work \cite{Schwinger}, it is known that the pair production is suppressed exponentially for a weak field $E$ due to the pair creation rate $\sim\exp(-\pi E_\mathrm{cr}/E)$. Therefore, for small values of polarization $\delta$, pair production manly occurs near the maximum of the electric field component $E_x(t)$ and the electric field component $E_y(t)$ plays the role in accelerating created particles. Because of the phase difference between $E_x(t)$ and $E_y(t)$, the particles created at the peak of the first half cycle of $E_x(t)$ are accelerated by $E_y(t)$ along the $y$ direction, while the particles created at the nadir of the second half cycle of $E_x(t)$ are accelerated by $E_y(t)$ along the $-y$ direction. As a consequence, the created particles at different positions will have different momentum component $q_y$, which results in two peaks of the final momentum distribution. This also explains why the distance of these two peaks increases with the value of polarization $\delta$ increasing. For large values of polarization, the particles created by $E_y(t)$ can also be split by the acceleration of $E_x(t)$, thus the previous two separated parts are connected again. Moreover, for a few-cycle field, pair production is dominated by the field component $E_x(t)$ at $t=0$, so there is no split of momentum distribution, see Fig. \ref{Fig1}. For a many-cycle field, however, near $t=0$ the electric field component $E_x(t)$ has several cycles having almost the same amplitude as $E_x(0)$, which induces the split of momentum distribution, see Figs. \ref{Fig2} and \ref{Fig3}.

Generally, EP pairs created from vacuum via tunneling process are considered to be at rest and then are accelerated by strong laser pulses. However, on the basis of the similarity between pair creation and strong-field ionization (cf. Fig. \ref{Fig3} here and Fig. 1 in Ref. \cite{PfeifferPRL2012}), the particles created from vacuum may have nonzero momentum. Of course, further studies are needed to prove this conjecture. Since many experiments about strong-field ionization are also carried out at $\gamma\sim\mathcal{O}(1)$, some explorations about nonperturbative multiphoton regime are useful. We know that in this intermediate regime both nonperturbative and perturbative process play an role in pair creation, so neither of them can fully explain the phenomena occurring in this regime. Thus a theory combining nonperturbative process with perturbative process will be hopeful to understand the experiment results \cite{PfeifferPRL2012,HofmannPRA2014}.

\begin{figure}[htbp]\suppressfloats
\includegraphics[width=8cm]{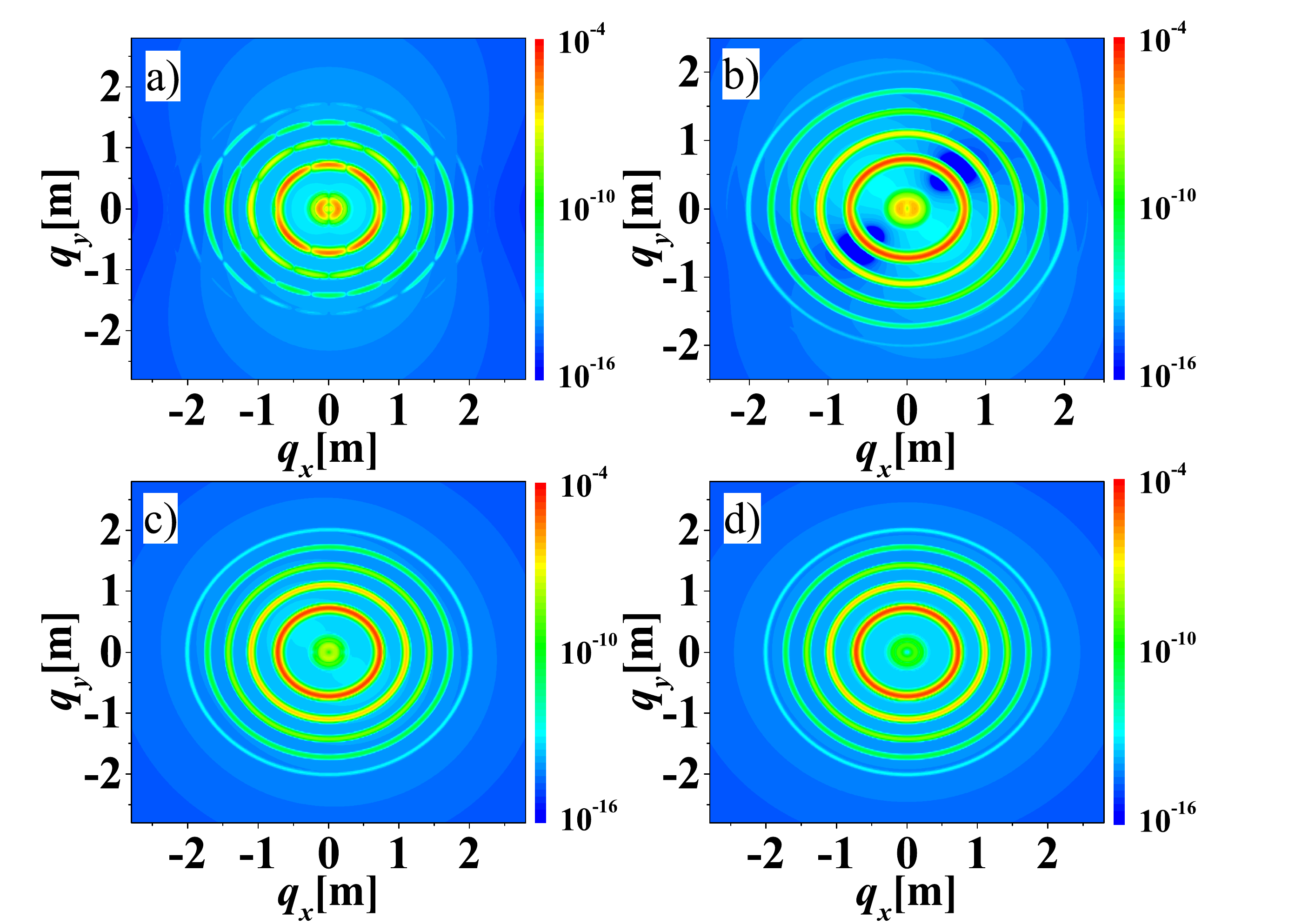}
\caption{\label{Fig4}(color online) Momentum spectra of created EP pairs in the $(q_x,q_y)$ plane for different polarizations with $\omega=0.5m$. (a) for $\delta=0$, (b) for $\delta=0.5$, (c) for $\delta=0.9$, and (d) for $\delta=1$. Other electric field parameters are the same as in Fig. \ref{Fig1}.}
\end{figure}

The momentum distributions of created EP pairs for different polarizations with $\gamma=3.54\sqrt{1+\delta^2}$ are shown in Fig. \ref{Fig4}. For the large laser frequency $\omega=0.5m$, one can clearly see some ring structures in the momentum spectra. These rings are formed by multiphoton absorbtion, i.e., absorbing $n$ photons of frequency $\omega$ to conquer the energy gap and produce EP pairs. From the inside out, the rings correspond to $5$-, $6$-, $7$-, $8$-, and $9$-photon absorbtion, respectively. For the linearly polarized field ($\delta=0$), we can calculate the ring radius of $n$-photon absorbtion $q_n=[(n\omega/2)^2-m^2-(m^2E_0/\omega E_{cr})^2/2]^{1/2}$ by employing the energy conservation equation $n\omega=2\sqrt{m^2_*+q_n^2}$ with the effective mass $m_*=m[1+(mE_0/\omega E_{cr})^2/2]^{1/2}$ \cite{Kohlfurst2014}. For instance, for $n=5$, the ring radius $q_5\approx0.7228m$ is in excellent agreement with the numerical result $q_{num}\approx0.7227m$. Since we keep the laser intensity of the electric field (\ref{eq1}) constant, the effective mass expression is approximately valid for arbitrary polarized fields according to the definition of effective mass $m_*=m[1+m^2\langle- A_\mu A^\mu\rangle/E_{\mathrm{cr}}^2]^{1/2}$ with four-vector $A_\mu$ \cite{Kohlfurst2014,Volkov}. So the ring radius can be roughly calculated by the expression $q_n$ for arbitrary polarized fields. For instance, the analytical result of the ring radius of $5$-photon absorbtion in Fig. \ref{Fig4}(c) $q_5\approx0.7228m$ is approximately identical to the numerical result $q_{num}\approx0.7267m$. The small difference between these two results is caused by the effect of polarization which cannot be considered by the effective mass model. It is also found that with the value of polarization increasing the rings become more uniform and the created pairs with small momentum decreases due to the rotation of the electric field.

\begin{figure}[htbp]\suppressfloats
\includegraphics[width=8cm]{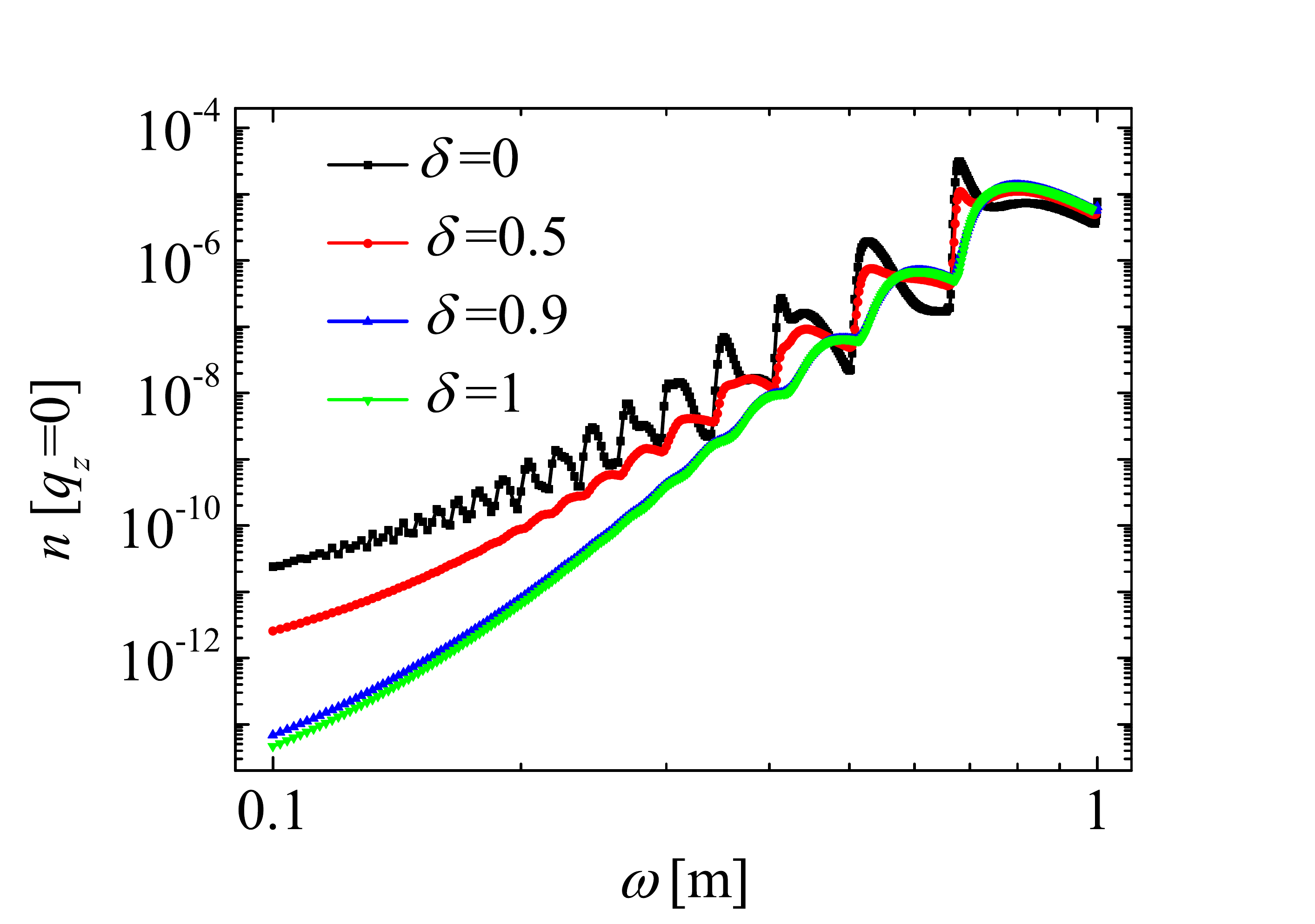}
\caption{\label{Fig5}(color online) The number density of created particles $n[q_z=0]$ as a function of the field frequency $\omega$ for $\delta=0$, $0.5$, $0.9$, and $1$, respectively. Other electric field parameters are the same as in Fig. \ref{Fig1}.}
\end{figure}

\begin{figure}[htbp]\suppressfloats
\includegraphics[width=8cm]{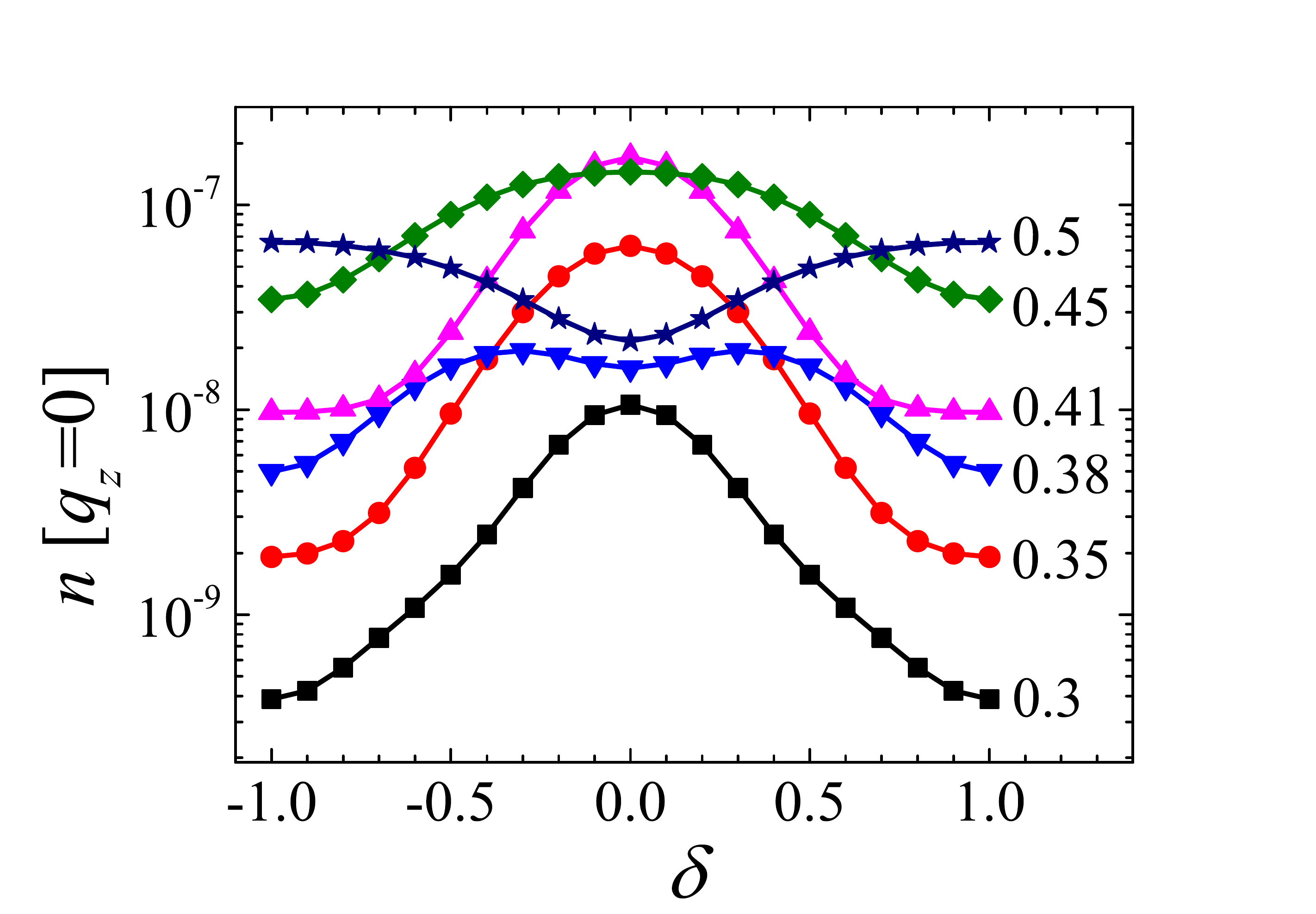}
\caption{\label{Fig6}(color online) The number density of created particles $n[q_z=0]$ as a function of the polarization $\delta$ for different field frequencies. Other electric field parameters are the same as in Fig. \ref{Fig1}.}
\end{figure}

From Figs. \ref{Fig1}-\ref{Fig3}, we can find that the maximum value of the momentum distribution function $f(+\infty)$ decreases with increasing the value of polarization. So one might speculate that the number density of created EP pairs may also have the same change tendency. However, things are more complicated than we originally thought. To figure out the relation between the number density and field polarization, we plot the number density scanning over the field frequency for different polarizations in Fig. \ref{Fig5}. It is found that there are many obvious oscillations on the curves and the ones corresponding to small field frequency vanish with the polarization increasing. These oscillations are caused by multiphoton pair production and have been studied for a linearly polarized field in Refs. \cite{Abdukerim2013,Kohlfurst2014}. Additionally, we also find that for a small field frequency the number density of created particles decreases greatly with the increase of the value of polarization. This is because for small laser frequencies nonperturbative Schwinger pair production is more significant. As pair creation is suppressed exponentially for weak fields, the number density is dominated by the electric field component with a large amplitude. Therefore, the polarized field with a small value of $\delta$, which has a larger amplitude of field component than the one with a large value of $\delta$, gives a high number density. For large field frequencies, we find that the relation between the number density and the polarization of electric field becomes more complicated because perturbative multiphoton process becomes more important. To see clearly, we plot the number density as a function of polarization for different laser frequency in Fig. \ref{Fig6}. It can be seen clearly that the number density of created pairs can reach its maximum at $\delta=0$, $\delta=\pm1$, or even $0<|\delta|<1$ for different field frequencies. Thus, for a large laser frequency the relation between the particle number density and the polarization of electric field is sensitive to the field frequency.

\section{Conclusion and discussion}
\label{Conc-Diss}

In summary, we have investigated the effects of electric field polarizations on vacuum pair production numerically by employing the real-time DHW formalism. First, we consider the effect of polarization on the momentum spectra of created particles and find that for a multi-cycle field the momentum distribution is split into two segments in the $q_y$ direction and then connected into a ring with increasing the polarization. This result is similar to the one in  SFI. By studying the effect of polarization on the number density of created EP pairs, we find that for a small laser frequency the particle number density deceases with the increase of field polarization. However, for a large laser frequency, the number density can reach its maximum value at linear polarization, circular polarization, or even elliptical polarization, which seems to be counterintuitive.

Due to the use of a general polarized field, our studies not only reveal several  characteristics of pair production which can not be captured by a linearly polarized or circularly polarized field, such as the split of momentum spectrum, but also link up pair production with the similar phenomena in SFI that some experiments are ongoing. Furthermore, because of the limitation in experiment instruments, a perfect circularly polarized field is harder to be produced than a elliptically polarized one. So our studies about the elliptic polarization are also more closely associated to the experiment interests.

\begin{acknowledgements}
This work was supported by the National Natural Science Foundation of China (NSFC) under Grant Nos. 11475026, 11175023 and 11335013, and also supported partially by the Open Fund of National Laboratory of Science and Technology on Computational Physics at IAPCM and the Fundamental Research Funds for the Central Universities (FRFCU).
\end{acknowledgements}

\end{document}